\documentstyle[aps,prl,multicol,epsfig]{revtex}
\newcommand{\be}{\begin{equation}}
\newcommand{\ee}{\end{equation}}
\newcommand{\bea}{\begin{eqnarray}}
\newcommand{\eea}{\end{eqnarray}}

\newcommand{\ep} {\epsilon} 
\begin{document}
\draft
\title{Coulomb ``blockade'' of Nuclear Spin Relaxation in Quantum Dots}
\author{
 Y. B. 
Lyanda-Geller$^{1,2}$
I.L. Aleiner$^3$, and B.L. Altshuler$^{4,5}$} 
\address{
$^1$Naval Research Laboratory, Washington DC 20375\\
$^2$Beckman Institute and Department of Physics
University of Illinois, Urbana, Illinois 61801\\
$^3$Department of Physics and Astronomy, SUNY at Stony Brook,
Stony-Brook, NY 11794\\
$^4$Department of Physics, Princeton University, Princeton, NJ 08544\\
$^5$NEC Research Institute, Princeton, NJ 08540}
\date{\today}
\maketitle
\begin{abstract}
We study the mechanism of 
nuclear spin relaxation in
quantum dots due to the electron exchange with 2D gas.  
We show that the nuclear spin relaxation
rate $T_1^{-1}$ is dramatically affected by the Coulomb blockade (CB)
and can be controlled by gate voltage. 
In the case of strong spin-orbit (SO) coupling the relaxation rate is
maximal in the CB valleys whereas for the weak SO
coupling  the maximum of $1/T_1$ is near the CB peaks. 
\end{abstract}
\pacs{PACS No: 76.60-k, 73.23Hk, 73.21La}
\begin{multicols}{2}
\narrowtext

Discreteness of energy spectra in quantum dots (QD), 
in combination with the Coulomb
electron-electron interactions, determines all their physical 
properties~\cite{REF:Aleiner}. 
In this paper, we study the impact of these two crucial features 
of QD on the interaction of electrons with nuclei.
In particular, we evaluate nuclear spin relaxation (NSR) rate, $1/T_1$, 
which determines the effectiveness of 
dynamical nuclear spin polarization~\cite{Meier} as well.
 
What mechanisms of NSR are efficient in the QD?  
Because even weak magnetic fields suppress the NSR  
due to the dipole-dipole interaction of nuclear spins~\cite{Meier},
we will consider NSR caused by the hyperfine coupling (HC) 
between nuclear, ${\bf S}$, 
and electron, $\mbox{\boldmath $\sigma$}$ spins:
\begin{equation}
{\cal H}_{h}=AV\sum\limits_{i}{\bf S}_i{\hat {\cal B}}({\bf r}_i),\quad 
{\hat {\cal B}}({\bf r}_i)=\Psi_{\alpha}^{+}({\bf r}_i)
\mbox{\boldmath $\sigma$}_{\alpha\beta}\Psi_{\beta}({\bf r}_i),
\label{HC}
\end{equation}
where $A$ is the HC strength, and $V$ is the crystal cell volume. 
%HC results in average 
Substitution of the operator $\hat{{\cal B}}$ by its
mean value, ${\bf {\cal B}}$, leads to the Hamiltonian of the
system of nuclear spins, ${\bf S}_i$ in 
an effective magnetic field, ${\bf {\cal B}}$. 
Being uniform, this field 
causes 
only a uniform precession of ${\bf S}_i$, 
rather than NSR.
Thermal fluctuations of ${\bf {\cal B}}$ in metals 
lead to the Korringa NSR mechanism~\cite{REF:Korringa}
 - each nuclear spin flip is accompanied by the 
creation of a {\it triplet} electron-hole pair.

In QD this process violates the
energy conservation:
the electron spectrum is discrete and 
Kramers degeneracy is lifted by the Overhauser nuclear field 
and/or external magnetic field. 
In order to overcome the mismatch of electron and nuclear spin splittings,  
a level broadening $\Gamma_0$ was included for electrons localized on 
donors~\cite{REF:Abragam,REF:Perel,REF:Paget} 
and in the quantum Hall effect 
regime~\cite{REF:Klitzing,REF:Vagner,REF:Macdonald}. 
However, $\Gamma_0$ must be introduced with 
a great caution. 
Indeed, any broadening implies that
the final state of the system is not 
a discrete 
eigenstate of the QD 
but rather 
an eigenstate of a larger system with a continuous
spectrum which includes QD as its subsystem. Thus, one has to specify the nature 
of the QD coupling to the outside world. 

We propose that charge exchange between the 
QD and the reservoirs (with continuous spectrum) leads to NSR.
Our main result is that NSR can be
controlled and changed by orders of magnitude
by the gate voltage, $V_g$, which defines the average number
of the electrons in the QD.
The origin of this effect is the Coulomb blockade
(CB)~\cite{REF:Averin_Likharev}
 that determines the tunneling of electrons
in and out of the QD.
Furthermore, 
$1/T_1$ in QD is governed not only by the 
QD electrons tunneling rates and the effective
fields ${\hat {\cal B}}({\bf r}_i)$,
but also by the probabilities for the QD to have a 
particular integer charge, $Q$, as well as by
the occupation numbers of the electronic  
states in the QD and in the leads at a given $Q$.
Moreover, we find that the NSR rate depends on the
sensitivity of the QD spectrum to the change in 
the nuclear spin configuration, which  
is maximal in the presence of electron spin-orbit coupling
(SOC).

Let us now discuss the role that SOC plays in NSR. 
Due to SOC the total spin of nuclei and electrons is not conserved. 
Each electron level, even if Kramers degeneracy is lifted, 
contains mixture of up and down spin states, 
and energy conservation law does not prevent nuclear spin flips 
(nuclear spin splitting $\hbar\omega\to 0$). 
Does the presence of SOC
leads to NSR even in absolutely closed dots?
The answer to this question is negative for the following reason.
Due to SOC, both the magnitude, and the direction of the
effective magnetic field, ${\bf {\cal B}}=<\hat{{\cal B}}>$,
become spatially inhomogenous but  
{\it remain time-independent}. 

Although the field originating from SOC, ${\bf {\cal B}}={\bf {\cal B}_{so}}({\bf r})$,
causes an inhomogeneous spin precession 
and thus broadens the NMR line, 
it can not lead to the actual NSR. 
The reason is that ${\bf {\cal B}_{so}}({\bf r})$
is time-independent.
One can separate this inhomogeneous spin precession 
by the spin echo technique~\cite{REF:Abragam}.
Furthermore, a time-independent ${\bf {\cal B}}$ 
can reduce the nuclear polarization but is unable to
eliminate it altogether. For example, 
if local nuclear polarization is parallel to ${\bf {\cal B}}$,
it will not change at all. 

However,  
even weak electron tunneling on and off the dot 
causes true NSR. 
The reason is that 
${\bf {\cal B}}_{so}$ is determined by the  
particular electronic configuration, and is 
modified by each tunneling event, 
thus becoming {\it time-dependent}. 
This dynamics leads to NSR,
because {\it it is not reversible}. 
Indeed, due to inhomogeneous precession of 
the nuclear spins, 
the adiabatic electron eigenstates change between tunneling events, 
and it is improbable that the system returns to the same electron/nuclear 
configuration. 

Let us now roughly estimate the 
NSR rate 
due to this mechanism, given the effective electron escape 
rate $\gamma$.
Typical rotation angle of 
${\bf S}$ within the time $\gamma^{-1}$ is 
${\cal B}_{so}/\gamma\ll 1$. 
At larger time scale, $t$, nuclear spins undergo random spin 
diffusion. 
The mean square angle deviation is 
$\langle \Theta^2(t)\rangle\sim ({\cal B}_{so}/\Gamma )^2 t\gamma =
{\cal B}_{so}^2 t/\gamma$.
Defining NSR rate $1/T_1$ by $\langle \Theta^2(T_1)\rangle\sim 1$, we 
obtain
$1/T_1\sim {\cal B}_{so}^2/\gamma$. The rate $\gamma$ 
can be fine tuned by adjusting the gate voltage, $V_g$, 
applied to the QD. 
$\gamma$ reaches maximum at the CB peak, where the dot conductance $G$ is 
maximal, and is minimal in the CB valley. 
%at minimal $G$. 
We reach a counterintuitive conclusion: 
NSR rate peaks
in the CB valleys, while near the CB peaks
it is suppressed, see Fig.2. As we shall see,
this conclusion holds as long as SOC is not too weak. 

The described physical situation is not unique for NSR in QD.
In particular, it resembles Mandelstam-Leontovich-Pollak-Geballe relaxation 
mechanisms of absorption of infrared radiation~\cite{REF:relaxational}: 
As long as the relaxation rate 
$\gamma_r$ exceeds the frequency of the radiation $\omega$, 
the relaxation rate is 
inverse proportional to $\gamma_r$. 
One can also recall the motional narrowing 
of spectral lines~\cite{REF:Anderson}.

Our goal is the microscopic theory of NSR in QD which includes 
the calculation of $\gamma$ and ${\cal B}_{so}$.
We start with the standard Hamiltonian of a 
QD connected to leads
(see e.g. Ref.~\cite{REF:Aleiner}
for the discussion of its validity)
$\hat{H}= \hat{H}_D +\hat{H}_{L} + \hat{H}_{LD}$, where 
$\hat{H}_D(t)$, $\hat{H}_L$ and $\hat{H}_{LD}$ describe, 
respectively,  the dot, the leads and tunneling between them:
\begin{equation}
\hat{H}_D = \sum\nolimits_{\alpha\alpha}
\epsilon_{\alpha}
a^{\dagger}_{\alpha}a_{\alpha}
+ E_c \left( \hat{n} - {\cal N}\right)^2,
\label{h1}
\end{equation}
$\hat{H}_L = \sum_{k,j}
\epsilon_{k,j}b^{\dagger}_{k,j} b_{k,j}, 
\hat{H}_{LD}=\sum_{k,j;\alpha}t_{\alpha}^j
b^{\dagger}_{k,j}a_{\alpha} + h.c.$. 
Here $a_{\alpha},a^{\dagger}_{\alpha}$ and 
$b_{k,j,},b^{\dagger}_{k,j}$ are creation/annihilation  
electron operators describing single-electron QD states $\alpha$,
and electrons in state $k$ in the lead $j$, respectively; 
$\hat{n}=\sum_{\alpha}a^{\dagger}_{\alpha}a_{\alpha}$ is the operator of 
the number of particles, ${\cal N}=CV_g/e$, 
$C$ is the capacitance of the dot, 
$e$ is the electron charge, and
$E_c=e^2/2C$ is the single-electron charging energy; 
matrix elements $t_{\alpha}^j$ determine the tunneling widths 
of QD states $|\alpha\rangle$, $\Gamma_t^{(\alpha)}= 
\sum_j 2\pi|t_{\alpha}^j|^2\nu$,
where $\nu$ is the density of states in leads. 

The NSR rate $1/T_1$ due to HC is 
connected to the linear
response of the electronic system to Zeeman magnetic field 
induced by nuclei at their location~\cite{REF:Abragam}:
\be
T_1^{-1} = 2A^2V^2T{\rm Im}\chi_{xx}(\omega)/\omega\vert_{\omega\to 0}.
\label{EQ:NSR}
\ee
The transverse spin susceptibility
$\chi_{xx}$ is defined by the response 
$\delta \langle s_x({\bf r})\rangle_{\omega}
=\chi_{xx}(\omega)h_\omega$, $h_\omega=\int dt\exp{(i\omega t)}h(t)$, 
of the electron spin density, $s_x({\bf r})$, to the perturbation 
\be 
\delta H=h(t) \sum\nolimits_{\alpha\beta}[s_x({\bf r})]_{\alpha\beta}
a^\dagger_\alpha a_\beta.
\label{perturb}
\ee 
As soon as SOC is included,
the diagonal matrix elements
$[s_x({\bf r})]_{\alpha\alpha}$, become finite and result in the 
shifts of the energy levels $\epsilon_{\alpha}$.
We assume the dot to be weakly coupled to leads which are 
in thermodynamic equilibrium.
Then the probability,  $P_j$, for the QD electrons to be in the state $j$, 
tends to follow 
the Gibbs distribution $P_j^G$ determined by
the instantaneous value of $\hat{H}(t)$. 
However, 
%because the  relaxation rate is finite, 
$P_j$ is delayed as compared with $\hat{H}(t)$
due to finite relaxation:
\bea
P_j(t) &=& P_j^G\left\{\hat{H}(t)\right\} \label{Mandelstam}\\
&-& \sum_k \left[\int_0^\infty dt_1
e^{-\hat{\Gamma}t_1}\right]_{jk} 
\frac{d}{dt} P_k^G\left\{\hat{H}(t)\right\} + \dots,
\nonumber
\eea 
%where the 
%relaxation 
The matrix $\hat\Gamma$ characterizes the 
%decay
relaxation 
of the deviation of $P_j$ from their equilibrium values:
\be
\frac{d}{dt} P_j(t)=- \sum_k \Gamma_{jk}
 P_k, \quad \sum_j \Gamma_{jk} =
\sum_k \Gamma_{jk}
 P_k^G=0.
\label{kinetics}
\ee
The second term in the RHS of Eq.~(\ref{Mandelstam}) gives rise to
the relaxational dissipation~\cite{REF:relaxational}.
The average spin equals to
\be
\langle s_x({\bf r})\rangle = \sum\nolimits_{j\alpha} n_\alpha^j P_j
s_{\alpha},\quad 
s_{\alpha}\equiv[s_x({\bf r})]_{\alpha\alpha}
\label{avspin}
\ee
where $n_\alpha^j=0,1$ is the occupation number of one-electron level
$\alpha$ for a given state of the electron system $j$.
\begin{figure}
\vskip-0.3truecm
%\vskip-4.3truecm
\epsfig{figure=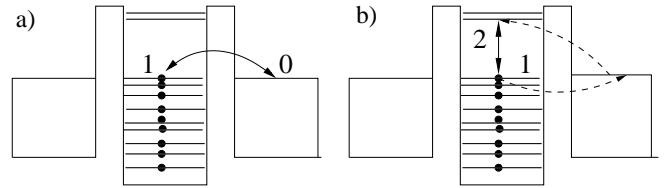,width=3.4in,rheight=2.8in,angle=0,silent=}    
 %\vskip-0.2truecm
 \vskip-4.5truecm
\caption{Processes dominating NSR in the (a) 
vicinity of and (b) away from  the Coulomb 
blockade peaks. 
Lines denote single-electron levels. Dots indicate filled levels. Arrows
show the relevant transition. Dashed lines in (b) show the actual electron 
processes leading to the relevant transition.}
\label{Fig1}
\end{figure}     
\vskip-0.3cm
\noindent 
Substituting $P_j(t)$ into Eqs.~(\ref{avspin}) and Eq.~(\ref{EQ:NSR}), we  
obtain $\chi_{xx}(\omega)$ and $T_1^{-1}$:
\be
\frac{1}{T_1} = - A^2V^2{T}\sum_{\alpha,\beta,j}
s_{\alpha}
 \left[\int_0^\infty dt_1
e^{-\hat{\Gamma}t_1}\right]_{\alpha j}
\frac{\partial P^G_j}
{\partial \epsilon_\beta}
s_{\beta}. 
\label{EQ:NSR2}
\ee
We emphasize that this NSR rate is
{\em inversely} proportional to the rate of the population relaxation of the
QD states, provided the latter rate is larger than $\omega$. 

As follows from Eq.~(\ref{EQ:NSR2}), the 
NSR is dominated by electron configurations
such that (i) their equilibrium probabilities are most sensitive to the
shifts of one-electron levels, and (ii) their 
relaxation is slowest. 
Let us identify the optimal configurations for
temperatures, $T$, below the mean level spacing, $\Delta$.
%, $T$ is the temperature, 
%and $\Delta $ is the mean level spacing. 
We begin with $V_g$ tuned to a CB peak. 
The energy distance to the peak
$U({\cal N})=E_c({\cal N}- N_0) \simeq \Delta$, 
with $N_0$ being a half-integer closest
to ${\cal N}$. Then the  relaxation occurs (see Fig.~\ref{Fig1}a) via
the two-level system (TLS) of the QD states with electron 
occupation numbers $N_0 \pm 1/2$, which we denote $1$ and $0$ respectively. 
Occupation of other states is suppressed exponentially, 
i.e. $P_0 + P_1 = 1$. 
The Gibbs distribution for such TLS 
gives $P_1^G/P_0^G = \exp\left[-(U({\cal N}) + \epsilon_1)/T\right]$
where $\epsilon_1$ is the energy of the filled electron level
in state ``1''.
The relaxation matrix for the two states is determined
by single tunneling width $\Gamma_{t}^{(1)}$
\be
\hat{\Gamma} = \Gamma_{t}^{(1)}
\pmatrix{1-f & -f \cr -1+f & f},
\label{g1}
\ee
where $f= \{1+\exp[(U({\cal N})+\ep_1)/T]\}^{-1}$ is the electron occupation 
number in the leads. From Eq.~(\ref{EQ:NSR2}), we find
\be
T_1^{-1} = \frac{A^2V^2|[s({\bf r})_x]_{11}|^2}{4\hbar\Gamma_{t}}
\cosh^{-2}\frac{|U({\cal N})|+\ep_1}{2T}.
\label{EQ:NSR3}
\label{peaks}
\ee
Thus, NSR decreases rapidly as $V_g$ deviates  from
the CB peaks, because 
the sensitivity of $P_j$ to $\epsilon_\alpha$
is exponentially suppressed by lifting the charge degeneracy.  
%%%%%%%%%%%%%%%%%%%%%%%%%%%%%%%%%%%%%%%%%%%%%%%%%%%%%%%

Naively, one may expect the Eq.~(\ref{EQ:NSR3}) to hold, at
least qualitatively, even far from the peaks.
%(we take $U(N) < 0$,
%the opposite case is considered analogously).
However, as soon as
$|U({\cal N})|$ exceeds $\Delta$,
another TLS becomes optimal, see
Fig.~\ref{Fig1}b. The energy difference between the two
configurations in this TLS (we will call them $1$ and $2$)
is $\epsilon_{eh}$ -- the smallest energy of the electron-hole
pair excited in the dot.
At $|U({\cal N})| \gtrsim \Delta$,  $\epsilon_{eh} < |U({\cal N})|$,
%this pair of states becomes the most probable one, and 
the population of all other states 
%can be neglected, 
is negligible. As a result, $P_1 + P_2=1$.
In the equilibrium,
$P_2^G/P_1^G=\exp\left[-(\epsilon_2(t)-\epsilon_1(t))/T\right]$, {\em
i.e.} in contrast with the previous case 
%it does not contain 
$P_{1,2}$ are $|U(N)|$-independent because
{\it now} the relevant excitation does not change the 
%number of particles in 
charge of the QD. 
Charging energy, however, determines 
the rate of the transitions between the two states. 
If $V_g$ is
not too far from the peak, 
$U({\cal N}) \lesssim T \ln (E_C^2/\Gamma_T \Delta)$, the 
transition $1\to 2$ proceeds
via the state $0$ (i.e. due to electron transfer between the QD and
the reservoir) as an intermediate real state. One finds
\begin{mathletters}
\be
\hat{\Gamma} =
\frac{\Gamma_{t}^{(1)}\Gamma_{t}^{(2)}e^{-\frac{|U({\cal N}|)}{T}}}
{\Gamma_{t}^{(1)}+\Gamma_{t}^{(2)}}
\pmatrix{ e^{\frac{\ep_1}{T}};
 & - e^{\frac{\ep_2}{T}}
\cr - e^{\frac{\ep_1}{T}}; & e^{\frac{\ep_2}{T}} }.
\label{2real}
\ee
This result can be derived from the master equation including
the state $0$.
%To understand its structure, let us discuss in more detail
For its interpretation consider the rate of the $1\to 2$ transition,
$[\hat{\Gamma}]_{11}$.
% characterizing $1\to 2$ transition.
%The factor  $\Gamma_{t}^{(1)}e^{-\frac{|U({\cal
%N})|+\ep_1}{T}}$ characterizes the inverse time of t
The electron escape rate out of the state $1$ is
$\Gamma_{t}^{(1)}e^{-\frac{|U({\cal N})|+\ep_1}{T}}$. 
After the electron leaves the QD, another one
enters during the time $\simeq 1/\Gamma_t$. 
%The probability that it
It occupies the state $2$ with the probability
$\Gamma_{t}^{(2)}/(\Gamma_{t}^{(1)}+\Gamma_{t}^{(2)})$.
The product of those two factors gives the probability of the
transition $1\to 2$.

At larger deviations 
%of $V_g$ 
from the CB peak, 
$|U({\cal N})| \gtrsim T \ln (E_c^2/\Gamma_t \Delta)$, 
processes changing the QD charge can be neglected.
%number of particle are exponentially suppressed. 
The relaxation is now determined by the
inelastic co-tunneling mechanism~\cite{REF:Averin,REF:Aleiner},
%which involve 
i.e., the state $0$ is used as a virtual state. 
%Instead of
The Eq.~(\ref{2real}) becomes
\be
\hat{\Gamma} =
\frac{\Gamma_{t}^{(1)}\Gamma_{t}^{(2)}\ep_{21}} {4 \pi \left[U({\cal
N})\right]^2}
\left[
\pmatrix{F -1 & -F -1 \cr -F+1 & F+1}
\right]
,
\label{2virtual}
\ee
where $\ep_{21}=\ep_2-\ep_1>0$, is the lowest energy of
the electron-hole excitation in the QD, $F=\coth (\ep_{21}/2T)$.
The factors in parenthesis are determined by the phase volume for
the electron-hole pair created in the reservoir as the result of
the inelastic co-tunneling process.
\end{mathletters}

\begin{figure}
\vskip-0.4truecm
\epsfig{figure=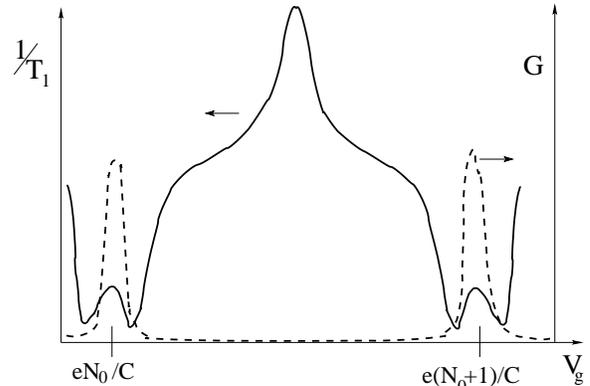,width=3.0in,rheight=1.7in,angle=0,silent=}    
 \vskip0.9truecm
\caption{Sketch of the predicted NSR dependence on  
$V_g$ in the presence of SOC. Dashed
line shows the linear conductance through the dot.}
\label{Fig2}
\vskip-0.3cm
\noindent
\end{figure}

Using the equilibrium 
%probabilities of 
occupation numbers of the two QD states
and Eqs.~(\ref{2virtual}) and~(\ref{EQ:NSR2}), we find
\bea
T_1^{-1} &=& \frac{A^2V^2\vert[s({\bf r})_x]_{11}-[s({\bf r})_x]_{22}\vert^2}
{4\hbar\ {\rm max}\ (\Gamma_t, \Gamma_c)
\cosh^2\frac{\ep_{21}}{2T}}
\label{valley} \\
\Gamma_t&=&\frac{2 \Gamma_{t}^{(1)}\Gamma_{t}^{(2)}
\cosh \frac{\ep_{21}}{2T}
} { \Gamma_{t}^{(1)}+\Gamma_{t}^{(2)}}e^{-\frac{|U({\cal N})|}{T}}
;\nonumber\\
\Gamma_c&=&\frac{\Gamma_{t}^{(1)}\Gamma_{t}^{(2)}\ep_{21}
\coth \frac{\ep_{21}}{2T}
} {2 \pi \left[U({\cal
N})\right]^2}.\nonumber
\eea
This result qualitatively differs from
Eq.~(\ref{EQ:NSR3}), and the dependence of $1/T_1$ on $V_g$
is non-trivial, see Fig.~\ref{Fig2}.
%%%%%%%%%%%%%%%%%%%%%%%%%%%%%%%%%%%%%%%%%%%%%%%%%%%%%%%

With yet further change of $V_g$ towards the bottom of a CB valley, 
one has to take into account the state $0$ with
an electron removed from the dot (hole-like process) and with
an electron added to the dot (electron-like process).
This changes the rates $\Gamma_{t,c}$ in Eq.~(\ref{valley}):
\begin{eqnarray}
\Gamma_t&=&\frac{4 \Gamma_{t}^{(1)}\Gamma_{t}^{(2)}
\cosh \frac{\ep_{21}}{2T} e^{-E_C/T}
} { \Gamma_{t}^{(1)}+\Gamma_{t}^{(2)}}
\cosh\frac{|U({\cal N})|-E_C}{T}
;\nonumber\\
\Gamma_c&=&\frac{\Gamma_{t}^{(1)}\Gamma_{t}^{(2)}\ep_{21}
\coth \frac{\ep_{21}}{2T}
} {2 \pi}
 \left[
\frac{1}
{|U({\cal N})|}
- \frac{1}
{2E_C-|U({\cal N})|}
\right]^2.
\label{deepvalley} 
\end{eqnarray}
Note that the inelastic rate 
$\Gamma_c$ vanishes at the bottom of the valley, 
because the electron- and
hole-like processes with the same final states are coherent, and their
amplitudes are opposite in sign due to the Fermi statistics
(the higher levels would give non-zero
contribution, but the conclusion
about the maximum of $1/T_1$ at the bottom of the CB valley
remains valid). 
Overall dependence of $1/T_1$ on $V_g$ in the presence of SOC is sketched on 
Fig.~\ref{Fig2}.

We now discuss NSR when SOC of electrons in QD is absent, and 
each QD level corresponds 
to a spin projection aligned or antialigned with magnetic field.
For typical distance between the lowest unoccupied QD level level 
and highest occupied QD level with opposite spin,
$\epsilon_2-\epsilon_1 \sim \Delta \gg \Gamma^{(i)}, T$, the calculation of 
the susceptibility $\chi_{xx}(\omega)$ 
is equivalent to the evaluation of the 
Fermi golden rule probability of 
the simultaneous electron and nuclear spin flip. 
In this process, the initial and final electron states are real, one or 
both of them are in the reservoir,
and energy-nonconserving electron spin flip inside the QD is
incorporated as the virtual transition. 

\begin{figure}
%\vskip-3.7truecm
\epsfig{figure=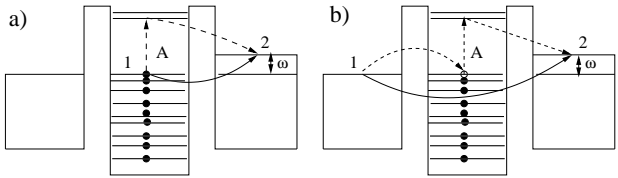,width=3.2in,rheight=2.8in,angle=0,silent=}    
%\vskip-0.3truecm
\vskip-4.5truecm
\caption{Dominant NSR processes at weak SOC
in the (a) vicinity of and (b) away from  the CB peaks 
($\epsilon_2-\epsilon_1 \gg \Gamma^{(i)}$). 
Solid lines with arrows denote transitions between initial 
and final states, dashed lines with arrows correspond to stages of 
those transitions. The spin flips occur in transitions ``A''.} 
\label{Fig3}
\end{figure}

In the vicinity of the CB peaks, 
the two relevant QD states are electron on the lower level (1) and 
electron in the leads (2), see Fig.~\ref{Fig3}. 
Deeper in the valleys the main contribution 
is due to the process analogous to the elastic cotunneling in the 
CB valley conductance (Fig.~\ref{Fig3}b): the initial and final QD 
electron configurations coincide but the electron-hole triplet 
pair with is excited in the leads.
\vskip-0.1cm
\noindent 
The result for the overall NSR rate dependence on $V_g$ in the absence of SOC is 
\be
\frac{1}{T_1}=
\frac{A^2 V^2|[s({\bf r})_x]_{12}|^2\Gamma_t^{(2)}
\left(\epsilon_{1}-\epsilon_2\right)^{-2}}
{ 2\hbar{\rm \min}\left(
\cosh^{-2}\frac{|U({\cal N})|+\ep_1}{2T},\
\frac { \pi \left[U({\cal N})\right]^2} {\Gamma_{t}^{(1)}T}\right)}
\label{nosoc}
\ee
Such NSR rate~\cite{anomclose} is {\it directly} proportional to the rate of the population 
relaxation of the QD states, is maximal at CB peaks, minimal in the valley, 
and much smaller than $1/T_1$ in the presence of SOC.

We now estimate the NSR rate. In GaAs, $A\!=40\mu$eV\cite{REF:Paget},
and $V\!=\!45\AA^3$. We consider QD formed in a 
50\AA -wide quantum well, 
with $0.1\mu$m in-plane dimensions, charging energy $E_c=0.5$meV, level spacing $\Delta=50\mu$eV, 
and tunneling width $\Gamma_t\!=\!2\mu$eV, at $T=100$mK. The typical matrix 
element $s_{11}$ is of the order of the inverse volume of the dot. Therefore 
we use $V^2s_{11}^2=0.8\times 10^{-12}$ in Eq.~(\ref{EQ:NSR3}), 
and obtain $T_1^{-1}= 0.25$Hz at the CB peak, and $T_1^{-1}= 2\times 10^{-3}$Hz 
in the NSR minimum close to the CB peak, at $U({\cal N})\!=\!55\mu$eV, Fig.2. 
In the CB valley, 
by using Eq.~(\ref{valley},\ref{deepvalley}), and assuming $s_{11}\!-\!s_{22}\sim 0.1s_{11}$, 
at $U({\cal N})=0.495$meV we obtain $T_1^{-1}=1.6\times 10^3$Hz. 
Our analysis shows also that even for QD with rather weak SOC, $T_1^{-1}$ 
in the CB valley has at least local maximum, so that the minimal 
NSR rate occurs at $V_g$ values between the CB peak and the CB valley bottom.
Thus, $V_g$ indeed changes $T_1^{-1}$ by orders of magnitude.

In conclusion, we have shown that nuclear spin relaxation 
in QD is strongly affected by the Coulomb blockade, as well as the 
spin-orbit coupling, and proposed the relaxational mechanism of NSR. 
The NSR rate is predicted to have non-trivial dependence
on $V_g$. Similarly to nuclei in experiments on quantum point 
contacts~\cite{REF:Wald}, nuclei shall affect transport properties of QD.    
The gate voltage depedence of NSR can be used, 
in particular, for creating and sustaining nuclear spin polarization in 
QD for electron spin filtering (we are grateful to C.M. 
Marcus for discussions of this experimental setting). 
This work was supported by DARPA QUIST program (Y.L.-G.) and (B.A.) and 
by Packard Foundation (I.A.).

\end{multicols}
\end{document}